\documentstyle[aaspp4,flushrt,11pt]{article}
\tighten

\begin{document}

\twocolumn

\title{Tests of Relativistic Gravity using Millisecond Pulsars}

\author{Jon Bell\altaffilmark{1}}

\slugcomment{To appear in: {\em Pulsar Timing, General Relativity, and the
Internal Structure of Neutron Stars}, Proceedings of a Colloquium held at
the Royal Netherlands Academy of Arts and Sciences, 24--28 September, 1996.}

\altaffiltext{1}{The University of Manchester, NRAL, Jodrell Bank,
Cheshire SK11~9DL, UK.~~~ Email: jb@jb.man.ac.uk}

\begin{abstract}
General relativity asserts that: energy and momentum conservation laws are
valid, preferred frames do not exist, and the strong equivalence principle is
obeyed. In this paper recent progress in testing these important principles
using millisecond pulsars is summarised.
\end{abstract}

\section{Introduction}

\def\bb{$\bullet$~}
The fundamental physics and principles that can be observed and
tested by the exceptional precision of pulsar timing includes
(\cite{bel97}):

\noindent
\bb Relativistic precession \\
\bb Shapiro delay \\
\bb Einstein delay \\
\bb Gravitational waves \\
\bb Variation in G \\
\bb Chandrasekhar mass \\
\bb Spin-orbit coupling \\
\bb Ultra low frequency gravitational waves \\
\bb Strong equivalence principle \\
\bb Lorentz Invariance  \\
\bb Conservation laws 

At this meeting Esposito-Far\`{e}se gave an update on the first 4 items and
summary of the parametrised post-Newtonian formalism (PPN).  Will (1993)
\nocite{wil93} also discusses the PPN formalism in detail and gives
limits on many of the ten PPN parameters.  Taylor et al. (1992)
\nocite{twdw92} discuss many other relativistic effects which could in
principle be measured with sufficient precision. Limits on the PPN
parameters $\alpha_1$, $\alpha_2$, $\alpha_3$ and $\xi$ will be discussed
here in relation to the last two items. Tests of the strong equivalence
principle (SEP) giving limits on $\Delta$ will also be discussed due to the
similar nature of the tests. These tests are null tests and it is the
90\% confidence level limits which are quoted.

In placing such limits, one wishes to know the extent to which strong field
effects are contributing. Measurement of a given PPN parameter
$\hat{\alpha}$ contains both a weak field contribution $\alpha $ and a
strong field contribution $\alpha^{'}$ (\cite{de92a})
\begin{equation}
\hat{\alpha} = \alpha + \alpha^{'}(c_1 + c_2 + \cdot\cdot\cdot) +
\cdot\cdot\cdot  
\end{equation}
Here $c_1$, $c_2$ represent the compactness ($E_{grav} /mc^2$) of the bodies
involved. For the sun, $c_i \sim 10^{-6}$, for a neutron star $c_i \sim
0.2$ and for a black hole $c_i \sim 0.5$. Therefore, strong field effects
are poorly constrained by solar system experiments, while pulsars provide
comparable sensitivity and ease of study when compared to black holes.

The cosmic microwave background (CMB) has been chosen as the absolute frame
in most studies. While some recent results (\cite{lp94}) have questioned
this, it is the magnitude, not the direction of the absolute
velocity ${\bf w}$ that is most relevant; this is similar for both the CMB
and Lauer \& Postman data.

One might ask whether the similar nature (i.e. upper limits from
low eccentricity orbits) of the tests discussed below (which constrain
$\Delta$, $\alpha_1$ and $\alpha_3$) makes them degenerate. This is not the
case; there are sufficient degrees of freedom and different figures of merit
for each test so that different pulsars are being used for each test.

\section{Lorentz Invariance, {\boldmath $|\alpha_2| < 2.4 \times 10^{-7}$}}
\label{a2}

If a gravitational interaction is not Lorentz invariant (PPN $\alpha_2 \ne
0$, due to some long-range tensor field), an oblate spinning body will feel
a torque (\cite{nor87b}):
\begin{equation}
{\bf \tau} \propto \alpha_2 {\bf w} \times {\bf \Omega},
\label{e;a2}
\end{equation}
where {\boldmath $\Omega$} is the angular velocity. This torque would cause
the spin axis to precess about ${\bf w}$. Since the spin-orbit coupling
between the sun and planets is weak, the close alignment ($\sim
6^{\circ}$) of the spin and orbital angular momenta means that the above
torque is weak. Quantitatively, the limit is $|\alpha_2| < 2.4 \times
10^{-7}$, showing that the gravitational interaction is Lorentz invariant to
high precision (\cite{nor87b}).

There are two important assumptions made here: primordial alignment of the
spin and orbital angular momenta and that the sun has not made many
rotations and by chance is closely aligned at the present epoch. Pulsars
play a role in confirming that the second assumption is valid since if the
torque was sufficiently large to cause the sun to make many rotations it
would also be large enough to cause the fastest pulsars to precess out of
view (\cite{nor87b}), assuming they do not have fan beams.

Nordtvedt (1987) \nocite{nor87b} also considered preferred location effects as
distinct from the above preferred frame effects. The resulting Lagrangian
for a three-body interaction contains the PPN parameter $\xi$. Using the
Galactic center as a distant third body yields  ${\bf \tau} \propto
\xi {\bf w} \times {\bf \Omega}$, giving a limit on $\xi$ similar
to the limit on $\alpha_2$ by the same arguments.

\section{Polarised orbits and Relativistic Precession}
\label{s;pre}

The tests discussed in Sections \ref{sep}, \ref{a1} and \ref{abp} search for
the presence of eccentricities induced in pulsar orbits. These are
gravitational analogues of the Stark effect, with the orbits being polarised
in particular directions. However there is a non-zero probability that the
relativistic precession of the orbit may cause cancellation with the
intrinsic eccentricity of the system.

The problem of the possible cancellation was first considered by Damour and
Sch\"afer (1991)\nocite{ds91}. They noted that, if the binary pulsar system
is old compared to the time scale for precession, so that many rotations had
been completed, then a statistical treatment of the probability of
cancellation was sufficient since the goal was an upper limit rather than a
measurement. A more precise statistical treatment was derived by Wex (1996)
\nocite{wex97} who also demonstrated the power of using multiple systems to 
improve the limits.

\section{Strong Equivalence Principle, {\boldmath $|\Delta| < 0.004$}}
\label{sep}

The SEP requires the universality of the free fall of self-gravitating
objects, i.e. in the same gravitational potential, two bodies should feel
the same acceleration regardless of their mass, composition and density.
Nordtvedt (1968) \nocite{nor68a,nor68b} showed that if SEP did not hold for
the Earth-Moon-Sun system, the Moon's orbit would be eccentric and polarised
with the semi-major axis pointing towards the Sun. So began the now famous
lunar-laser-ranging experiments which searched for this polarisation using
the Apollo 11 reflector and measurement uncertainties of $\sim 1$ ns in the
time of flight.

Damour and Sch\"{a}fer (1991) \nocite{ds91} pointed out the need for such a
test in a strong field regime and showed that it is possible using binary
pulsars. They suggested that the Earth-Moon-Sun system be replaced with a
pulsar-companion-Galaxy system. If the companion is a white dwarf, the
composition, density and self gravity is very different to the pulsar giving
sensitivity to strong field effects. Damour and Sch\"{a}fer (1991)
\nocite{ds91} showed that the figure of merit for choosing the best test
systems is $f_{\Delta} = P_b^2/e$ and used PSR B1953+29 to obtain the limit
$|\Delta| = |1 - M_I/M_G| < 0.01$. Arzoumanian (1995) \nocite{arz95}
suggested that PSR B1800--27 could be used to improve this limit to
$|\Delta| < 0.004$, however it is not clear that this system is sufficiently
old (\cite{wex97}). If several pulsars are used simultaneously the
multiplication of small probabilities leads to the very rigorous bound of
$|\Delta| < 0.004$ (\cite{wex97}).

\section{Lorentz Invariance, {\boldmath $|\hat{\alpha}_1| < 1.7 \times 10^{-4}$}}
\label{a1}

If preferred reference frames exist and $\alpha_1 \ne 0$, then there is a
constant forcing term in the time evolution of the eccentricity vector of a
binary system. For a very low eccentricity orbit, this tends to
``polarize'' the orbit, aligning the eccentricity vector with the projection
onto the orbital plane of the absolute velocity of the system. Hence, the
orbital parameters of very low eccentricity binary pulsars such as PSR
B1855+09 may be used to set an upper bound of $|\hat{\alpha}_1| < 5 \times
10^{-4}$ (\cite{de92a}). This compares with limits from solar system data of
${\alpha}_1 = 2.1 \pm 1.9 \times 10^{-4}$ (\cite{hel84}). 

The most circular orbit known ($e \sim 10^{-6}$), that of PSR J2317+1439
(\cite{cnt96}) has a figure of merit $f_{\alpha_1} = P_b^{1/3}/e$ 10 times
better than PSR B1855+09. However, the more unfortunate orientation with
respect to the CMB and poorly constrained radial velocity means that only a
factor of 3 improvement was possible, giving a limit of $|\hat{\alpha}_1| <
1.7 \times 10^{-4}$ (\cite{bcd96}).

\section{Conservation Laws and Lorentz Invariance, {\boldmath $|\alpha_3| <
2.2 \times 10^{-20}$}} 

As shown by Nordtvedt and Will (1972) \nocite{nw72,wil93}, a non-zero
$\alpha_{3}$ induces a contribution to the perihelion precession of the
planets in the solar system. The two planets with the best measurements of
periastron advance were Earth and Mercury. By combining the observations for
two planets it is possible to eliminate the terms involving other
parameters, obtaining $|49 \alpha_{1} -
\alpha_{2} - 6.3 \times 10^{5} \alpha_{3} - 2.2\xi| < 0.1$
(\cite{wil93}). Using the limits on $\alpha_1$, $\alpha_2$, $\xi$ a limit of
$|\alpha_{3}| < 2 \times 10^{-7}$ was thus obtained. 

\subsection{Single Pulsars}

A tighter limit on $\alpha_{3}$ has been obtained by considering the effect
of the acceleration  
\begin{equation}
{\bf a}_{self} \propto \alpha_3 {\bf w} \times {\bf \Omega}
\label{e;a3}
\end{equation}
on the observed pulse periods of isolated pulsars.  The observed pulse
period $P \simeq P_{0}(1 + v_{r}/c)$, contains a contribution from the Doppler
effect due to the radial velocity $v_{r}$.  Similarly any radial
acceleration $a_{r}$ contributes to the observed period derivative $\dot{P}
\simeq \dot{P}_{0} + P a_{r} / c$.

Self accelerations are directed perpendicular to both ${\bf w}$ and ${\bf
\Omega}$.  If self accelerations were contributing strongly to the observed
period derivatives of pulsars, roughly equal numbers of positive and
negative observed period derivatives would be expected, since the spin axes
and therefore the self accelerations are randomly oriented.  The observed
distribution of normal pulsars (excluding those pulsars in globular
clusters) contains only positive period derivatives, allowing a limit of
$|\alpha_{3}| < 2 \times 10^{-10}$ to be placed (\cite{wil93}). Using
millisecond pulsars, Bell (1996) \nocite{bel96} obtained a limit of
$|\alpha_{3}| < 5 \times 10^{-16}$. 

\subsection{Binary Pulsars}
\label{abp}

For a binary pulsar with a white dwarf companion, we again have two bodies
with very different self gravities and sensitivities to strong field
effects. If $\alpha_3 \ne 0$, the induced self-acceleration of the white
dwarf would be negligible compared to that of the pulsar. Hence, we now have
a rocket in a binary system. Since ${\bf a}_{self}$ is perpendicular to
${\bf \Omega}$ and since the spin and orbital angular momenta are aligned
for recycled systems, the ${\bf a}_{self}$ is in the plane of the orbit. The
resulting effect is a polarised orbit similar to those predicted by SEP
violations (Section \ref{sep}) and Lorentz invariance violations (Section
\ref{a1}) (\cite{bd96}).

The figure of merit for choosing the best test systems is
$f_{\alpha_3} = P_b^2/eP$. Selecting appropriate systems and applying the
statistical treatment of relativistic precession (Section \ref{s;pre}) gives
a limit of $|\alpha_{3}| < 2.2 \times 10^{-20}$ (\cite{bd96}). The only
other ultra-high-precision null experiments (giving limits of order
$10^{-20}$ on a dimensionless theoretical parameter) of which we are aware,
are the recent Hughes-Drever-type tests (\cite{pre85,lam86,chu89}), shown in
Figure 14.2 of Will (1993)\nocite{wil93}. It is remarkable that tests
involving binary pulsars can rank, together with modern laser-cooled trapped
atom experiments, among the most precise null experiments of physics.

\section{Prospects for Further Improvements}

\begin{figure}[h]
\plotone{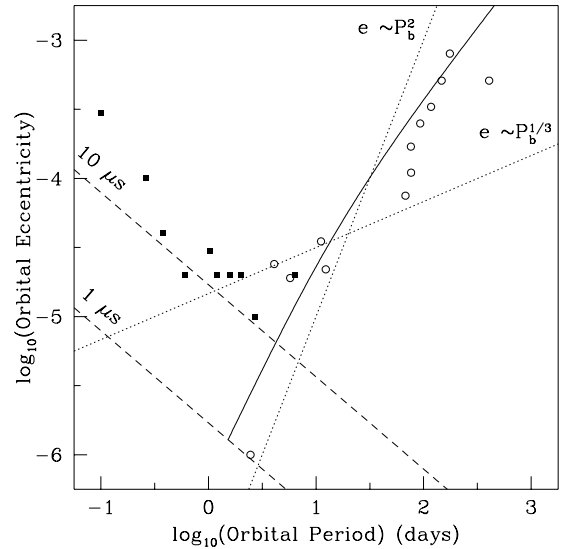}
\caption{Orbital eccentricities of radio pulsars.  Circles ---
low-mass binary pulsars. Dashed lines -- approximate limits obtainable on
eccentricities for rms timing residuals of $10\,\mu$s and $1\,\mu$s.}
\label{pbe}
\end{figure}

The figures of merit indicate how strongly these tests depend on the orbital
periods and eccentricities of binary pulsars. The dotted lines in Figure
\ref{pbe} indicate the relative slopes of $f_{\Delta} \propto P_b^2 /e$ and
$f_{\alpha_1} \propto P_b^{1/3} /e$. There are also strongly evolutionary
links expected between $P_b$ and $e$ (\cite{phi92b}) as shown by the solid
line. Comparison of the slope of this curve, with the figure of merit
dependence on $P_b$ and $e$ indicates that scope for improvements of the SEP
test is small unless more longer orbital period systems can be found. A
similar conclusion for the $\alpha_3$ limit can be drawn but the additional
dependence on $P$, ($f_{\alpha_3} \propto P_b^2 /eP$) makes it less clear.

The flatter dependence on $P_b$ of $f_{\alpha_1}$ means that short orbital
period systems are preferable. However, only upper limits are presently
available for many of these (Figure \ref{pbe}). If these upper limits could
be reduced substantially, it should be possible to usefully improve the
limit on $\alpha_1$, especially if several systems are used.

The limits on $\alpha_2$ and $\xi$ may be improved slightly, by careful
consideration of pulse profile changes of the fastest pulsars to obtain
limits on the precession. However there is another more promising
approach. If the orientation of the spin and orbital angular momenta could
be determined for binary millisecond pulsars it would be possible to improve
the limits on $\alpha_2$ and $\xi$ by several orders of magnitude. The
inclination of the PSR J1012+5307 orbit has been determined from optical
observations (\cite{vbk96}). It may be possible to obtain the orientation of
the pulsar spin axis from polarisation observations. With only one system,
the ambiguity of many rotations would remain remain, but with several such
binary systems, statistical arguments similar to those used for $\Delta$ and
$\alpha_3$ could provide very strict limits.

\acknowledgements
I thank the organizers for an excellent meeting and G.
Esposito-Far\`{e}se for useful discussions.





\end{document}